\documentclass[ aip, jmp, reprint,]{revtex4}
\usepackage{amsfonts}
\usepackage{url}
\usepackage{amssymb}
\usepackage{amsmath}
\usepackage{graphicx}
\usepackage{dcolumn}
\usepackage{bm}

\begin{document}

\title[]{\textbf{Absolute stability of axisymmetric perturbations in
strongly-magnetized \\
collisionless axisymmetric accretion disk plasmas}}
\author{C. Cremaschini}
\affiliation{International School for Advanced Studies (SISSA), Trieste, Italy}
\affiliation{Consortium for Magnetofluid Dynamics, University of Trieste, Trieste, Italy}
\author{J. C. Miller}
\affiliation{Department of Physics (Astrophysics), University of Oxford, Oxford, UK}
\affiliation{International School for Advanced Studies (SISSA), Trieste, Italy}
\author{M. Tessarotto}
\affiliation{Department of Mathematics and Informatics, University of Trieste, Trieste,
Italy}
\affiliation{Consortium for Magnetofluid Dynamics, University of Trieste, Trieste, Italy}
\date{\today }

\begin{abstract}
The physical mechanism responsible for driving accretion flows in
astrophysical accretion disks is commonly thought to be related to the
development of plasma instabilities and turbulence. A key question is
therefore the determination of consistent equilibrium configurations for
accretion-disk plasmas and investigation of their stability properties. In
the case of collisionless plasmas kinetic theory provides the appropriate
theoretical framework. This paper presents a kinetic description of
low-frequency and long-wavelength axisymmetric electromagnetic perturbations
in non-relativistic, strongly-magnetized and gravitationally-bound
axisymmetric accretion-disk plasmas in the collisionless regime. The
analysis, carried out within the framework of the Vlasov-Maxwell
description, relies on stationary kinetic solutions of the Vlasov equation
which allow for the simultaneous treatment of non-uniform fluid fields,
stationary accretion flows and temperature anisotropies. It is demonstrated
that, for such solutions, no axisymmetric unstable perturbations can exist
occurring on characteristic time and space scales which are long compared
with the Larmor gyration time and radius. Hence, these stationary
configurations are actually stable against axisymmetric kinetic
instabilities of this type. As a fundamental consequence, this rules out the
possibility of having the axisymmetric magneto-rotational or thermal
instabilities to arise in these systems.
\end{abstract}

\maketitle



A fundamental issue in the physics of accretion disks (ADs) concerns the
stability of equilibrium or quasi-stationary configurations occurring in AD
plasmas. The observed transport phenomena giving rise to the accretion flow
are commonly ascribed to the existence of instabilities and the subsequent
development of fluid or MHD turbulence \cite%
{Balbus2003,Lomi09,Narayan2006,Rebusco}. In principle, these can include
both MHD phenomena (such as drift instabilities driven by gradients of the
fluid fields) and kinetic ones (due to velocity-space anisotropies,
including, for example, trapped-particle modes, cyclotron and Alfven waves,
etc.). Possible candidates for the angular momentum transport mechanism are
usually identified either with the magneto-rotational instability (MRI) \cite%
{Velikhov1959,Chandrasekhar1961,B-H1991} or the thermal instability (TMI)
\cite{Field1965,Shakura1973,Shakura1976,Mond}, caused by unfavorable
gradients of rotation/shear and temperature respectively. The validity of
the above identifications needs to be checked in this case, because they
usually rely on incomplete physical descriptions, which ignore the
microscopic (kinetic) plasma behavior. In fact, \textquotedblleft
stand-alone\textquotedblright\ fluid and MHD approaches which are not
explicitly based on kinetic theory and/or do not start from consistent
kinetic equilibria, may become inadequate or inapplicable for collisionless
or weakly-collisional plasmas. Apart from possible gyrokinetic and finite
Larmor-radius effects (which are typically not included for MRI and TMI),
this concerns consistent treatment of the kinetic constraints which must be
imposed on the fluid fields (see related discussion in Refs.\cite%
{Cr2010,Cr2011}). This concerns, in particular, the correct determination of
the constitutive equations for the relevant fluid fields. Because of this,
the issue of stability of these systems is in need of further study.

In this regard, some relevant background materials are provided by Refs.\cite%
{Cr2010,Cr2011,Cr2011a}, where a perturbative kinetic theory for
collisionless plasmas has been developed and the existence of asymptotic
kinetic equilibria has been demonstrated for axisymmetric magnetized
plasmas. In AD plasmas, in particular, they are characterized by the
presence of \emph{stationary azimuthal and poloidal species-dependent flows}
and can support \emph{stationary kinetic dynamo effects}, responsible for
the self-generation of azimuthal and poloidal magnetic fields \cite{CT},
together with \emph{stationary accretion flows}. This provides the basis for
a systematic stability analysis of such systems. We stress that these
features arise as part of the kinetic equilibrium solution, and are not
dependent on perturbative instabilities. Furthermore, by assumption in the
theory developed here there is no background (i.e., externally-produced)
radiation field. In principle, for a collisionless plasma at equilibrium,
charged particles can be still subject to EM radiation produced by
accelerating particles (EM radiation-reaction). However, the effect of these
physical mechanisms is negligible for the dynamics of non-relativistic
plasmas, and therefore they can be safely ignored in the present treatment.

In this paper we address the stability of these equilibria with respect to
infinitesimal axisymmetric perturbations. We restrict attention to the
treatment of non-relativistic, strongly-magnetized and gravitationally-bound
(see definition below) collisionless AD plasmas around compact objects for
which the theory developed in Refs.\cite{Cr2010,Cr2011} applies. The plasmas
can be considered \emph{quasi-neutral} and characterized by a \emph{%
mean-field interaction.} Accretion disks fulfilling these requirements rely
necessarily on kinetic theory in the so-called Vlasov-Maxwell statistical
description, which represents the fundamental physical approach for these
systems. In AD plasmas, electromagnetic (EM) fields can be present, which
may either be externally produced or self-generated. At equilibrium, they
are taken here to be axisymmetric and of the general form $\mathbf{B}%
^{\left( eq\right) }\equiv B^{\left( eq\right) }\mathbf{b}=\mathbf{B}%
_{T}^{\left( eq\right) }+\mathbf{B}_{P}^{\left( eq\right) }$ and $\mathbf{E}%
^{\left( eq\right) }\equiv -\nabla \Phi ^{\left( eq\right) }\left( \mathbf{x}%
\right) $. Here $\mathbf{B}_{T}^{\left( eq\right) }\equiv I(\mathbf{x}%
)\nabla \varphi $ and $\mathbf{B}_{P}^{\left( eq\right) }\equiv \nabla \psi (%
\mathbf{x})\times \nabla \varphi $ are the toroidal and poloidal components
of the magnetic field respectively, with $I(\mathbf{x})$ and $\Phi ^{\left(
eq\right) }\left( \mathbf{x}\right) $ being the toroidal current and the
electrostatic potential. Furthermore, $(R,\varphi ,z)$ denote a set of
cylindrical coordinates, with $\mathbf{x}=\left( R,z\right) $, while $\left(
\psi ,\varphi ,\vartheta \right) $ is a set of local magnetic coordinates,
with $\psi $ being the so-called poloidal flux function. The validity of the
previous representation for\ $\mathbf{B}^{\left( eq\right) }$ requires the
existence of locally nested magnetic $\psi -$surfaces, represented by $\psi
=const.$, while the expressions for $\psi (\mathbf{x})$, $I(\mathbf{x})$ and
$\Phi ^{\left( eq\right) }\left( \mathbf{x}\right) $ follow from the
stationary Maxwell equations. The gravitational field is treated here
non-relativistically, by means of the gravitational potential $\Phi
_{G}=\Phi _{G}(\mathbf{x})$. This means that the electrostatic and
gravitational fields are formally replaced by the effective electric field $%
\mathbf{E}_{s}^{eff}=-\nabla \Phi _{s}^{eff}$, determined in terms of the
effective electrostatic potential $\Phi _{s}^{eff}=\Phi ^{\left( eq\right) }(%
\mathbf{x})+\frac{M_{s}}{Z_{s}e}\Phi _{G}(\mathbf{x})$, with $M_{s}$ and $%
Z_{s}e$ denoting the mass and charge, respectively, of the $s$-species
particle (where $s$ can indicate either ions or electrons). Based on
astronomical observations, the magnetic field magnitudes are expected to
range in the interval $B\sim 10^{1}-10^{8}G$ \cite{Narayan,Frank,Vietri}.
This implies that the proton Larmor radius $r_{Li}$ is in the range $%
10^{-6}-10^{3}cm$ (the lower values corresponding to the lower temperature
and the higher magnetic field). Additional important physical parameters are
related to the species number density and temperature. Astrophysical AD
plasmas can have a wide range of values for the particle number density $%
n_{s}$, depending on the circumstances considered. Here we focus on the case
of collisionless and non-relativistic AD plasmas assuming values of the
number density $n_{s}$ in the range $n_{s}\sim 10^{6}-10^{15}cm^{-3}$. For
reference, the highest value of this interval corresponds to ion mass
density $\rho _{i}\sim 10^{-9}gcm^{-3}$. The choice of this parameter
interval lies well in the range of values which can be estimated for the
so-called radiatively inefficient accretion flows (RIAFs, \cite{Narayan,Tsi}%
). For these systems, estimates for species temperatures usually lie in the
ranges $T_{i}\sim 1-10^{5}keV$ and $T_{e}\sim 1-10keV$ for ions and
electrons respectively. Depending on the magnitude of the EM, gravitational
and fluid fields, the AD plasmas can sustain a variety of notable physical
phenomena, the systematic treatment of which requires their classification
in terms of\ suitable dimensionless parameters. These are identified with $%
\varepsilon _{M,s}$, $\varepsilon _{s}$ and $\sigma _{s}$, to be referred to
as \textit{Larmor-radius}, \textit{canonical momentum} and \textit{%
total-energy parameters}. Their definitions are respectively: $\varepsilon
_{M,s}\equiv \frac{r_{Ls}}{\left( \Delta L\right) ^{eq}},$ $\varepsilon
_{s}\equiv \left\vert \frac{M_{s}Rv_{\varphi }}{\frac{Z_{s}e}{c}\psi }%
\right\vert $ and $\sigma _{s}\equiv \left\vert \frac{\frac{M_{s}}{2}v^{2}}{{%
Z_{s}e}\Phi _{s}^{eff}}\right\vert $. Here, $r_{Ls}\equiv v_{ths}/\Omega
_{cs}$ denotes the Larmor radius of the species $s$, $v_{ths}$ and $\Omega
_{cs}$ are the species thermal velocity and the Larmor frequency
respectively, $\left( \Delta L\right) ^{eq}$ is the characteristic
scale-length of the equilibrium fluid fields, $\mathbf{v}$ is the
single-particle velocity and $v_{\varphi }\equiv \mathbf{v}\cdot R\nabla
\varphi $. Systems satisfying the asymptotic ordering $0\leq \sigma
_{s},\varepsilon _{s},\varepsilon ,\varepsilon _{M,s}\ll 1$ are referred to
as \emph{strongly-magnetized and gravitationally-bound} plasmas \cite%
{Cr2010,Cr2011}, with the parameters $\sigma _{s},\varepsilon _{s}$ and $%
\varepsilon _{M,s}$ to be considered as independent while $\varepsilon
\equiv \max \left\{ \varepsilon _{s},s=1,n\right\} $. In the following, we
shall assume that the poloidal flux is of the form $\psi \equiv \frac{1}{%
\varepsilon }\overline{\psi }(\mathbf{x}),$ with $\overline{\psi }(\mathbf{x}%
)\sim O(\varepsilon ^{0})$, while the equilibrium electric field satisfies
the constraint $\frac{\mathbf{E}^{\left( eq\right) }\cdot \mathbf{B}^{\left(
eq\right) }}{\left\vert \mathbf{E}^{\left( eq\right) }\right\vert \left\vert
\mathbf{B}^{\left( eq\right) }\right\vert }\sim O\left( \varepsilon \right) $%
. This implies that to leading-order $\Phi ^{\left( eq\right) }$ is a
function of $\psi $ only, while $\Phi _{s}^{eff}$ remains generally a
function of the type $\Phi _{s}^{eff}=\overline{\Phi }_{s}^{eff}(\overline{%
\psi },\vartheta )$ (see Ref.\cite{Cr2011}). At equilibrium, by
construction, the particle toroidal canonical momentum $p_{\varphi s}\equiv
\frac{Z_{s}e}{c}\psi _{\ast s}=M_{s}Rv_{\varphi }+\frac{Z_{s}e}{c}\psi $,
the total particle energy $E_{s}\equiv Z_{s}e\Phi _{\ast s}=\frac{M_{s}}{2}%
v^{2}\mathbf{+}{Z_{s}e}\Phi _{s}^{eff}$ and the magnetic moment $%
m_{s}^{\prime }$ predicted by gyrokinetic theory are either exact or
adiabatic invariants. In particular, the above orderings imply the
leading-order asymptotic perturbative expansions for the variables $\psi
_{\ast s}$ and $\Phi _{\ast s}$:%
\begin{equation}
\psi _{\ast s}=\psi \left[ 1+O\left( \varepsilon _{s}\right) \right] ,
\label{psi}
\end{equation}%
\begin{equation}
\Phi _{\ast s}=\Phi _{s}^{eff}\left[ 1+O\left( \sigma _{s}\right) \right] ,
\label{es}
\end{equation}%
while similarly $m_{s}^{\prime }=\frac{M_{s}w^{\prime 2}}{2B^{\prime }}\left[
1+O\left( \varepsilon _{M,s}\right) \right] $. From here on, we will use the
notation that primed quantities are always evaluated at the guiding-center.
In particular, $\mathbf{w}^{\prime }=\mathbf{v}-u^{\prime }\mathbf{b}%
^{\prime }-\mathbf{V}_{eff}^{\prime }$ denotes the perpendicular particle
velocity in the local frame having the effective drift velocity $\mathbf{V}%
_{eff}^{\prime }\equiv \frac{c}{B^{\prime }}\mathbf{E}_{s}^{eff}\times
\mathbf{b}^{\prime }$, while $u^{\prime }\equiv \mathbf{v}\cdot \mathbf{b}%
^{\prime }$. In the following we shall also assume that the toroidal and
poloidal magnetic fields and the species accretion and azimuthal flow
velocities scale as $\frac{\left\vert \mathbf{B}_{T}\right\vert }{\left\vert
\mathbf{B}_{P}\right\vert }\sim O\left( \varepsilon \right) $ and $\frac{%
\left\vert \mathbf{V}_{accr,s}\right\vert }{\left\vert \mathbf{V}_{\varphi
,s}\right\vert }\sim O\left( \varepsilon \right) $ respectively.

In validity of the previous assumptions, an explicit asymptotic solution of
the Vlasov equation can be obtained for the kinetic distribution function $%
f_{\ast s}^{eq}$ (KDF). As pointed out in Ref.\cite{Cr2011}, ignoring
slow-time dependencies, this is of the generic form $f_{\ast s}^{eq}=f_{\ast
s}^{eq}\left( X_{\ast s},(\psi _{\ast s},\Phi _{\ast s})\right) $. Here $%
X_{\ast s}$ are the invariants $X_{\ast s}\equiv \left( E_{s},\psi _{\ast
s},p_{\varphi s}^{\prime },m_{s}^{\prime }\right) $, while the brackets $%
(\psi _{\ast s},\Phi _{\ast s})$ denote implicit dependencies for which the
perturbative expansions (\ref{psi}) and (\ref{es}) are performed. Therefore,
$f_{\ast s}^{eq}$ is by construction an adiabatic invariant, defined on a
subset of the phase-space $\Gamma =\Omega \times U$, with $\Omega \subset
\mathbb{R}
^{3}$ and $U\equiv
\mathbb{R}
^{3}$ being, respectively, a bounded subset of the Euclidean configuration
space and the velocity space. Hence, $f_{\ast s}$ varies slowly in time on
the slow-time-scale $\left( \Delta t\right) ^{eq}$, i.e. $\frac{d}{dt}\ln
f_{\ast s}^{eq}\sim \frac{1}{\left( \Delta t\right) ^{eq}}$. In view of the
previous orderings holding for AD plasmas, this implies also $\frac{\left(
\Delta t\right) ^{eq}}{\tau _{col,s}}\ll 1$, where $\tau _{col,s}$ denotes
the Spitzer collision time for the species $s$. Therefore, this requirement
is consistent with the assumption of a collisionless plasma. A possible
realization of $f_{\ast s}^{eq}$ is provided by a non-isotropic generalized
bi-Maxwellian KDF. As shown in Ref.\cite{Cr2011}, $f_{\ast s}^{eq}$
determined in this way describes Vlasov-Maxwell equilibria characterized by
quasi-neutral plasmas which exhibit species-dependent azimuthal and poloidal
flows as well as temperature and pressure anisotropies. The existence of
these equilibria is warranted by the validity of suitable kinetic
constraints (see the discussion in Ref.\cite{Cr2011,CT}). As a consequence,
the same equilibria are characterized by the presence of fluid fields
(number density, flow velocity, pressure tensor, etc.) which are generally
non-uniform on the $\psi -$surfaces.

Let us now pose the problem of linear stability for Vlasov-Maxwell
equilibria of this type. This can generally be set for perturbations of both
the EM field and the equilibrium KDF, which exhibit appropriate time and
space scales $\left\{ \left( \Delta t\right) ^{osc},\left( \Delta L\right)
^{osc}\right\} $. Here both are prescribed to have \emph{fast time} and
\emph{fast space} dependencies with respect to those of the equilibrium
quantities, in the sense that%
\begin{equation}
\frac{\left( \Delta t\right) ^{osc}}{\left( \Delta t\right) ^{eq}}\sim \frac{%
\left( \Delta L\right) ^{osc}}{\left( \Delta L\right) ^{eq}}\sim O(\lambda ),
\label{ORDERING -DELTA}
\end{equation}%
with $\lambda $ being a suitable infinitesimal parameter. In the case of
strongly-magnetized AD plasmas, to permit a direct comparison with the
literature, we also assume that these perturbations are \emph{non-gyrokinetic%
}. In other words, they are characterized by typical wave-frequencies and
wave-lengths which are much larger than the Larmor gyration frequency $%
\Omega _{cs}$ and radius $r_{Ls}$. This implies that the following
inequalities must hold:%
\begin{equation}
\frac{\tau _{Ls}}{\left( \Delta t\right) ^{osc}}\sim \frac{r_{Ls}}{\left(
\Delta L\right) ^{osc}}\ll 1,  \label{U-3}
\end{equation}%
with $\tau _{Ls}=1/\Omega _{cs}$, while $\lambda $ must satisfy $\lambda \gg
\sigma _{s},\varepsilon _{s},\varepsilon ,\varepsilon _{M,s}.$ These will be
referred to as \emph{low-frequency }and \emph{long-wavelength perturbations}
with respect to the corresponding Larmor scales.\textbf{\ }Notice that Eqs.(%
\ref{ORDERING -DELTA}) and (\ref{U-3}) are independent and complementary,
establishing the upper and lower limits for the range of magnitudes of both $%
\left( \Delta t\right) ^{osc}$ and $\left( \Delta L\right) ^{osc}$. We now
determine the generic form of the perturbations as implied by the above
assumptions. For this purpose, we shall require in the following that the EM
field is subject to \textit{axisymmetric EM perturbations} of the form $%
\delta \mathbf{B}=\nabla \times \delta \mathbf{A}$, $\delta \mathbf{E}%
=-\nabla \delta \phi -\frac{1}{c}\frac{\partial \delta \mathbf{A}}{\partial t%
}$, with $\left\{ \delta \phi \left( \frac{\overline{\psi }}{\lambda },\frac{%
\vartheta }{\lambda },\frac{t}{\lambda }\right) ,\delta \mathbf{A}\left(
\frac{\overline{\psi }}{\lambda },\frac{\vartheta }{\lambda },\frac{t}{%
\lambda }\right) \right\} $ both assumed to be \emph{analytic }(with respect
to $\overline{\psi }$ and $\vartheta $) and \emph{infinitesimal}, i.e., such
that $\frac{\delta \mathbf{E}}{\left\vert \mathbf{E}^{\left( eq\right)
}\right\vert },\frac{\delta \mathbf{B}}{\left\vert \mathbf{B}^{\left(
eq\right) }\right\vert }\sim O(\varepsilon )$. This implies that the
corresponding perturbations for the EM potentials must scale as $\frac{%
\delta \phi }{\left\vert \Phi ^{\left( eq\right) }\right\vert },\frac{\delta
\mathbf{A}}{\left\vert \mathbf{A}^{\left( eq\right) }\right\vert }\sim
O(\varepsilon )O(\lambda ),$ with $\mathbf{A}^{\left( eq\right) }$ denoting
the equilibrium vector potential. As a consequence%
\begin{equation}
\frac{d}{dt}E_{s}=q_{s}\left[ \frac{\partial \delta \phi }{\partial t}-\frac{%
1}{c}\mathbf{v}\cdot \frac{\partial \delta \mathbf{A}}{\partial t}\right] .
\label{energy}
\end{equation}%
Similarly, the perturbation of the equilibrium KDF is taken of the general
form%
\begin{equation}
\delta f_{s}\equiv \delta f_{s}\left( X_{\ast s},(\psi _{\ast s},\Phi _{\ast
s}),\frac{\overline{\psi }}{\lambda },\frac{\vartheta }{\lambda },\frac{t}{%
\lambda }\right) ,  \label{delta-f}
\end{equation}%
with $\frac{\delta f_{s}}{f_{\ast s}^{eq}}\sim O(\varepsilon )O(\lambda ).$
It follows that the corresponding KDF (the solution of the Vlasov kinetic
equation) must now be of the general form%
\begin{equation}
f_{s}=f_{s}\left( X_{\ast s},(\psi _{\ast s},\Phi _{\ast s}),\frac{\overline{%
\psi }}{\lambda },\frac{\vartheta }{\lambda },\frac{t}{\lambda }\right) ,
\label{FORM}
\end{equation}%
while, from the Maxwell equations, the perturbations $\left\{ \delta \phi
,\delta \mathbf{A}\right\} $ are necessarily linear functionals of $\delta
f_{s}$. However, for analytic perturbations of the form (\ref{FORM}), $f_{s}$
must itself be regarded as an analytic function of $\overline{\psi }$ and $%
\vartheta $. Therefore, invoking Eqs.(\ref{psi}) and (\ref{es}), the same
KDF can always be considered as an asymptotic approximation obtained by
Taylor expansion of a suitable \emph{generalized KDF} of the form $%
f_{s}^{(gen)}\equiv f_{s}^{(gen)}\left( X_{\ast s},\left( \psi _{\ast
s},\Phi _{\ast s},Y_{\ast s}\right) ,\frac{t}{\lambda }\right) ,$ with $%
Y_{\ast s}\equiv \left[ \frac{\varepsilon _{s}\psi _{\ast s}}{\lambda },%
\frac{\sigma _{s}\Phi _{\ast s}}{\lambda }\right] $. In particular, denoting
$\delta f_{s}^{(gen)}\equiv f_{s}^{(gen)}-f_{\ast s}^{eq}$, it follows that
also $\delta f_{s}^{(gen)}$ is such that $\delta f_{s}^{(gen)}\equiv \delta
f_{s}^{(gen)}\left( X_{\ast s},\left( \psi _{\ast s},\Phi _{\ast s},Y_{\ast
s}\right) ,\frac{t}{\lambda }\right) $. Then, by Taylor expansion with
respect to the variables $Y_{\ast s}$, the perturbation $\delta
f_{s}^{(gen)} $ can be shown to be related to $\delta f_{s}$ (defined by Eq.(%
\ref{delta-f})) by%
\begin{equation}
\delta f_{s}^{(gen)}\cong \delta \widehat{f}_{s}\left( X_{\ast s},(\psi
_{\ast s},\Phi _{\ast s}),\frac{\overline{\psi }}{\lambda },\frac{\vartheta
}{\lambda }\right) \exp ^{i\omega t},  \label{ddf-gen}
\end{equation}%
where corrections of $\frac{O\left( \varepsilon _{s}\right) }{O(\lambda )}$
and $\frac{O\left( \sigma _{s}\right) }{O(\lambda )}$ have been neglected.
Similarly, invoking again Eqs.(\ref{psi}) and (\ref{es}), for the analytic
perturbations $\left\{ \delta \phi ,\delta \mathbf{A}\right\} $ we can
introduce the corresponding \emph{generalized perturbations }$\left\{ \delta
\phi ^{\left( gen\right) },\delta \mathbf{A}^{\left( gen\right) }\right\} $.
Neglecting in the similar way corrections of $\frac{O\left( \varepsilon
_{s}\right) }{O(\lambda )}$ and $\frac{O\left( \sigma _{s}\right) }{%
O(\lambda )}$, these are given as follows:%
\begin{eqnarray}
\delta \phi ^{\left( gen\right) }\left( Y_{\ast s},\frac{t}{\lambda }\right)
&\cong &\delta \widehat{\phi }\left( \frac{\overline{\psi }}{\lambda },\frac{%
\vartheta }{\lambda }\right) \exp ^{i\omega t},  \label{delta-f-gen} \\
\delta \mathbf{A}^{\left( gen\right) }\left( Y_{\ast s},\frac{t}{\lambda }%
\right) &\cong &\delta \widehat{\mathbf{A}}\left( \frac{\overline{\psi }}{%
\lambda },\frac{\vartheta }{\lambda }\right) \exp ^{i\omega t}.
\label{delta-a-gen}
\end{eqnarray}%
Analogous expressions for the corresponding generalized perturbations can be
readily obtained. In particular, using Eq.(\ref{ddf-gen}), we get the
following representation for $\delta f_{s}^{(gen)}$:%
\begin{equation}
\delta f_{s}^{(gen)}=\delta \widehat{f}_{s}^{\left( gen\right) }\left(
X_{\ast s},\left( \psi _{\ast s},\Phi _{\ast s},Y_{\ast s}\right) \right)
\exp ^{i\omega t},  \label{delta-f-gen3}
\end{equation}%
where, expanding the Fourier coefficient and neglecting again corrections of
$\frac{O\left( \varepsilon _{s}\right) }{O(\lambda )}$ and $\frac{O\left(
\sigma _{s}\right) }{O(\lambda )}$, $\delta \widehat{f}_{s}^{\left(
gen\right) }\cong \delta \widehat{f}_{s}\left( X_{\ast s},(\psi _{\ast
s},\Phi _{\ast s}),\frac{\varepsilon \psi }{\lambda },\frac{\vartheta }{%
\lambda }\right) $. Therefore, in view of Eq.(\ref{energy}), for
infinitesimal axisymmetric analytical EM perturbations $\left\{ \delta \phi
,\delta \mathbf{A}\right\} $, to leading order in $\lambda $ the Vlasov
equation implies the dispersion equation%
\begin{eqnarray}
&&\left. -i\omega \delta \widehat{f}_{s}\left( X_{\ast s},(\psi _{\ast
s},\Phi _{\ast s}),\frac{\overline{\psi }}{\lambda },\frac{\vartheta }{%
\lambda }\right) =\right. \\
&&\left. =i\omega q_{s}\left[ \delta \widehat{\phi }\left( \frac{\overline{%
\psi }}{\lambda },\frac{\vartheta }{\lambda }\right) -\frac{1}{c}\mathbf{v}%
\cdot \delta \widehat{\mathbf{A}}\left( \frac{\overline{\psi }}{\lambda },%
\frac{\vartheta }{\lambda }\right) \right] \frac{\partial f_{\ast s}^{eq}}{%
\partial E_{s}}.\right.  \notag
\end{eqnarray}%
\ Apart from the trivial solution $\omega =0$ (i.e., a stationary
perturbation of the equilibrium), this requires that, for $\omega \neq 0$,
one must have%
\begin{equation}
\delta \widehat{f}_{s}=-q_{s}\left[ \delta \widehat{\phi }-\frac{1}{c}%
\mathbf{v}\cdot \delta \widehat{\mathbf{A}}\right] \frac{\partial
f_{s}^{(eq)}}{\partial E_{s}},  \label{final}
\end{equation}%
where, by construction, $\delta \widehat{f}_{s},$ $\delta \widehat{\phi }$
and $\delta \widehat{\mathbf{A}}$ are manifestly independent of $\omega $.
Hence, Eq.(\ref{final}) necessarily holds also when $\left\vert \omega
\right\vert $ is arbitrarily small. In this limit $\left\{ \delta \widehat{%
\phi },\delta \widehat{\mathbf{A}},\delta \widehat{f}_{s}\right\} $ tend
necessarily to infinitesimal stationary perturbations of the equilibrium
solutions. On the other hand, Eqs.(\ref{delta-f-gen}), (\ref{delta-a-gen})
and (\ref{delta-f-gen3}) show that $\left\{ \delta \widehat{\phi },\delta
\widehat{\mathbf{A}},\delta \widehat{f}_{s}\right\} $ are always
asymptotically close to the generalized quantities $\left\{ \delta \widehat{%
\phi }^{\left( gen\right) },\delta \widehat{\mathbf{A}}^{\left( gen\right)
},\delta \widehat{f}_{s}^{\left( gen\right) }\right\} $, which are by
definition equilibrium perturbations [i.e., functions of $\left( \frac{%
\varepsilon _{s}\psi _{\ast s}}{\lambda },\frac{\sigma _{s}\Phi _{\ast s}}{%
\lambda }\right) $]. Since the latter again represent an equilibrium and are
independent of $\omega $, it follows that the only admissible solution of
the dispersion equation (\ref{final}) is clearly independent of $\omega $ as
well and coincides with the null solution, i.e.%
\begin{eqnarray}
\delta \widehat{\phi }\left( \frac{\overline{\psi }}{\lambda },\frac{%
\vartheta }{\lambda }\right) &\equiv &0, \\
\delta \widehat{\mathbf{A}}\left( \frac{\overline{\psi }}{\lambda },\frac{%
\vartheta }{\lambda }\right) &\equiv &0, \\
\delta \widehat{f}_{s}\left( X_{\ast s},(\psi _{\ast s},\Phi _{\ast s}),%
\frac{\overline{\psi }}{\lambda },\frac{\vartheta }{\lambda }\right) &\equiv
&0.
\end{eqnarray}%
In summary: \emph{no analytic, low-frequency and long-wavelength
axisymmetric unstable perturbations can exist in non-relativistic
strongly-magnetized and gravitationally-bound axisymmetric collisionless AD
plasmas.} We stress that this result follows from two basic assumptions. The
first one is the requirement that the equilibrium magnetic field admits
locally nested $\psi -$surfaces. The second one is due to the assumed
property of AD plasmas to be gravitationally-bound. This implies (as pointed
out above) that the effective ES potential $\Phi _{s}^{eff}$ is necessarily
a function of both $\psi $ and $\vartheta $, and therefore the perturbation
of the KDF is actually close to a function of the exact and adiabatic
invariants $X_{\ast s}$. A notable aspect of the conclusion is that it
applies to collisionless Vlasov-Maxwell equilibria having, in principle,
arbitrary topology of the magnetic field lines which can belong to either
closed or open magnetic $\psi -$surfaces. Also, as pointed out in Refs.\cite%
{Cr2010,Cr2011}, for strongly-magnetized plasmas these equilibria can give
rise to kinetic dynamo effects simultaneously with having accretion flows.
These results are important for understanding the phenomenology of
collisionless AD plasmas of this type. In particular, they completely rule
out the possibility that axisymmetric perturbations, which are
long-wavelength and low-frequency in the sense of the inequalities (\ref{U-3}%
), could give rise to kinetic instabilities in such systems. This conclusion
applies for collisionless AD plasmas (having in particular particle
densities within the range mentioned earlier) which are strongly-magnetized
and simultaneously gravitationally-bound . Since fluid descriptions of these
plasmas can only be arrived at on the basis of the present Vlasov-Maxwell
statistical description, also MHD instabilities, such as the axisymmetric
MRI \cite{Lomi09,Quataert2002}, the axisymmetric TMI (see for example \cite%
{Shakura1973,Shakura1976,Mond}), and axisymmetric instabilities driven by
temperature anisotropy (e.g., the firehose instability \cite{Rosin}) remain
definitely forbidden for collisionless plasmas under these conditions.


\textbf{Acknowledgments - }This work has been partly developed in the
framework of MIUR (Italian Ministry of University and Research) PRIN
Research Programs and the Consortium for Magnetofluid Dynamics, Trieste,
Italy.


\bigskip

\begin{thebibliography}{99}
\bibitem{Balbus2003} S.A. Balbus, Annu. Rev. Astron. Astrophys. \textbf{41},
555 (2003).

\bibitem{Lomi09} A.B. Mikhailovskii, J.G. Lominadze, A.P. Churikov and V.D.
Pustovitov, Plasma Physics Reports \textbf{35}, 4, 273-314 (2009).

\bibitem{Narayan2006} B. Mukhopadhyay, N. Afshordi and R. Narayan, Advances
in Space Research \textbf{38}, 12, 2877-2879 (2006).

\bibitem{Rebusco} P. Rebusco, O.M. Umurhan, W. Kluzniak and O. Regev, Phys.
Fluids \textbf{21}, 076601 (2009).

\bibitem{Velikhov1959} E.P. Velikhov, J. Exptl. Theoret. Phys. \textbf{36},
1398 (1959).

\bibitem{Chandrasekhar1961} S. Chandrasekhar, Proc. Natl. Acad. Sci. \textbf{%
46}, 253 (1961).

\bibitem{B-H1991} S.A. Balbus and J.F. Hawley, ApJ \textbf{376}, 214 (1991).

\bibitem{Field1965} G.B. Field, Ap. J. \textbf{142}, 531 (1965).

\bibitem{Shakura1973} N.I. Shakura and R.A. Sunyaev, A\&A \textbf{24}, 337
(1973).

\bibitem{Shakura1976} N.I. Shakura and R.A. Sunyaev, MNRAS \textbf{175}, 613
(1976).

\bibitem{Mond} E. Liverts, M. Mond and V. Urpin, MNRAS \textbf{404}, 283
(2010).

\bibitem{Cr2010} C. Cremaschini, J.C. Miller and M. Tessarotto, Phys.
Plasmas \textbf{17}, 072902 (2010).

\bibitem{Cr2011} C. Cremaschini, J.C. Miller and M. Tessarotto, Phys.
Plasmas \textbf{18}, 062901 (2011).

\bibitem{Cr2011a} C. Cremaschini and M. Tessarotto, Phys. Plasmas \textbf{18}%
, 112502 (2011).

\bibitem{CT} C. Cremaschini, J.C. Miller and M. Tessarotto, Proc. of the
International Astronomical Union \textbf{6}, 228-231 (2010),
doi:10.1017/S1743921311006995.

\bibitem{Narayan} R. Narayan, R. Mahadevan and E. Quataert, \textit{Theory
of Black Hole Accretion Discs}, 148, ed. M. Abramowicz, G. Bjornsson and J.
Pringle, Cambridge University Press, Cambridge (UK) (1998).

\bibitem{Frank} J. Frank, A. King and D. Raine, \textit{Accretion power in
astrophysics}, Cambridge University Press, Cambridge (UK) (2002).

\bibitem{Vietri} M. Vietri, \textit{Foundations of High-Energy Astrophysics}%
, University Of Chicago Press, Chicago, USA (2008).

\bibitem{Tsi} D. Tsiklauri, New Astronomy \textbf{6}, 487 (2001).

\bibitem{Quataert2002} E. Quataert, W. Dorland and G.W. Hammett, Astrophys.
J. \textbf{577}, 524-533 (2002).

\bibitem{Rosin} M.S. Rosin, A. Schekochihin, F. Rincon and S.C. Cowley,
MNRAS \textbf{413}, 7-38 (2011).
\end{thebibliography}
\end{document}